\documentstyle[12pt]{article} 
\newcommand{\tr}{\mathop{\rm tr}\nolimits} 
\newcommand{\Ker}{\mathop{\rm Ker}\nolimits} 
\newcommand{\im}{\mathop{\rm Im}\nolimits} 
\newcommand{\mod}{\mathop{\rm \; mod \;}\nolimits} 

\begin{document} 
\title{GENERALIZED HEISENBERG MODEL}
\author{S.\,Yu.\,Gubanov\/\thanks{gubanov@itp.ac.ru} \\
{\it Landau Institute for Theoretical Physics}\\
{\it Chernogolovka, 142432, Russia}} 
%\date{}
\maketitle

\begin{abstract}
We consider the $XXZ$ model for a chain of particles whose spins are 
arbitrary with the anisotropy parameter equal to the root of minus 
one and generalized periodic boundary conditions. The conditions for
the truncation of the functional fusion relations of the transfer 
matrices are obtained. The truncation results in a closed system of 
equations whose solution allows obtaining the energy spectrum.
%\\
%\\
%PACS: 05.50.+q
\\
\\
published: {\it Theoretical and Mathematical Physics, 130(3): 383-390 (2002)}
Translated from Teoreticheskaya i Matematicheskaya Fizika,\\
Vol. 130, No. 3, pp. 451-459, March, 2002. \\
Original article submitted July 19, 2001; revised September 2, 2001.
\end{abstract}

\paragraph{Introduction.}
The anisotropic Heisenberg $XXZ$-model describes a chain of $N$ two-level 
atoms and has the Hamiltonian
\begin{equation}
H_{XXZ}= \sum_{i=1}^{N-1} \left( \sigma^{+}_i 
\sigma^{-}_{i+1}+\sigma^{-}_i \sigma^{+}_{i+1}+\frac{q+q^{-1}}{4} 
\; \sigma^z_i \sigma^z_{i+1} \right),
\end{equation}
where $\sigma^{\mu}_i$ are the Pauli matrices and $q$ is the anisotropy 
parameter. Hamiltonian acts in the Hilbert space of states 
$({\bf C}^2)^{\otimes N}$, 
whose dimensionality is $2^N$. In 1931, Bethe proposed a general method 
(i.e., the Bethe anzatz) that allows reducing the problem of obtaining 
the eigenvalues and eigenvectors of the Hamiltonian to a problem known 
from the classical calculus, namely, the problem of obtaining solutions 
of a certain system of transcendental equations called the Bethe 
equations~\cite{GA}. The connection established later between quantum 
one-dimensional models and classical two-dimensional models of statistical
physics showed that the central role in the Bethe method belongs to a 
relation known as the Yang-Baxter equation. The quantum inverse problem 
method proposed by Faddeev, Takhtadzhyan, and Sklyanin allowed systematizing
and developing these ideas in a more algebraic form~\cite{TF}. 
The two-dimensional statistical six-vertex model (the ice model) is 
connected with the quantum one-dimensional Heisenberg $XXZ$ model. 
The connection between the transfer matrix $T(u)$ of the ice model and the 
Hamiltonian of the Heisenberg $XXZ$ model
\begin{equation}
\label{SVYAZ}
H_{XXZ} = -\frac{N}{2} \cos(\eta) + \sin(\eta) 
\frac{d}{d u} \log T_{\hat{\frac12},\frac12}(u)|_{u=0},
\end{equation}
where $e^{i \eta}=q$, determines the correspondence between the energy 
eigenvalues and the transfer-matrix eigenvalues. The operators $T(u)$ 
and $H_{XXZ}$ commute and therefore have the same system of eigenvectors.
It was shown in~\cite{ABBBQ} and~\cite{PS}, which focused on the 
Heisenberg model with the anisotropy parameter equal to the root of 
minus one ($q^{p+1}=-1$), that if the $XXZ$-model is reduced onto a 
specially selected space of states, than the resulting new model is 
equivalent in the thermodynamic limit ($N \to \infty$) to a
certain minimal model of the conformal field theory. For example, in 
the case of open boundary conditions of a special form the central 
charge is $c=1-6/p(p+1)$. There exists an infinite family of 
transfer matrices. They commute with each other and also satisfy an
infinite system of recursive functional fusion relations. 
It was shown in~\cite{BelStr} that under certain conditions, the
fusion relations for the transfer matrices of the six-vertex model 
are truncated and become a closed system of equations on eigenvectors 
such that the reduction on these eigenvectors in the thermodynamic 
limit leads to the minimal models of the conformal field theory. 
In other words, studying the quantum group reduction of the Heisenberg 
model and studying the conditions for the truncation of the fusion 
relations for the transfer matrices of the six-vertex model are the 
same problem. An algebraic explanation for the truncation of the fusion 
relations was given in~\cite{We, BelGub}. That the truncation results 
in a closed system of equations that allows obtaining the spectrum 
without solving the traditional transcendental Bethe equations presents
additional intrinsic interest. The result can sometimes be obtained 
in an explicit form~\cite{BelGub}. The preceding discussion applies
only to the chains of particles whose spins are $\frac12$. In this 
work, we investigate the Heisenberg $XXZ$-model for a chain of 
particles with arbitrary spins with generalized periodic (twisted) 
boundary conditions. We first obtain the functional fusion relations 
for transfer matrices with an arbitrary value of spin. We then 
investigate the necessary conditions for the truncation of these
relations. For this, we use the method based on the second linearly 
independent solution of the Baxter equation proposed in~\cite{BelStr}, 
generalized to the case of an arbitrary spin. We discuss suficient 
conditions for the truncation that are obtained from an algebraic 
analysis of the transfer-matrix structure. The case of the
$XXZ$-chain of particles with spin $1$ is examined as an example.

\paragraph{Fusion relations for transfer matrices of the six-vertex model
with arbitrary spin.} 
Let the numbers $l$ and $j$ determine the respective spins in the quantum 
and auxiliary spaces. The transfer matrix $T_{l, j}(u)$ acts in the 
configuration (quantum) space $({\bf C}^{2l+1})^{\otimes N}$, 
commutes for different values of the spectral parameter $u$ and the 
auxiliary space spin $j$,
\begin{equation}
[T_{l, j_1}(u_1), T_{l, j_2}(u_2)]=0,
\end{equation}
and is therefore a generating function of the integrals of motion. 
In the six-vertex model by definition, we have
\begin{equation}
T_{l, j} (u) \; = \; \tr_{\pi_j} \left( q^{-2\beta \gamma^z}
R_{l \otimes j}(u)\ldots R_{l \otimes j} (u) \right),
\end{equation}
where $\beta$ is the twist parameter, i.e., the number characterizing 
the boundary conditions (the case where $\beta=0$ corresponds to the 
usual periodic conditions). The matrix elements of the operator $\gamma^z$
in the representation with the spin $j$ have the form
\begin{equation}
\pi_j(\gamma^z)_{n k} = (j+1-n)\delta_{n k}; \quad n,k=1,2,\ldots,2j+1.
\end{equation}
The operator $R_{l \otimes j}(u)$ acts in the space 
${\bf C}^{2l+1} \otimes {\bf C}^{2j+1}$, 
and is a solution of the Yang-Baxter equation
\begin{equation}
\label{YANGBAXTER}
R_{l \otimes j}(u-v) 
R_{l \otimes j'}(u) R_{j \otimes j'}(v) =
R_{j \otimes j'}(v) R_{l \otimes j'}(u)
R_{l \otimes j}(u-v).
\end{equation}
For example, in the case of the quantum spin $l=\frac12$, 
$R$-matrix of the six-vertex model has the form
\begin{equation}
\label{R22}
R_{\hat{\frac12} \otimes j} (u) =
\left(
\begin{array}{cccc}
\sin(u+(\frac{1}{2}+\hat{H})\eta) & \sin(\eta) \hat{F} \\
\sin(\eta) \hat{E} & \sin(u+(\frac{1}{2}-\hat{H})\eta) \\
\end{array}
\right),
\end{equation}
where the operators $\hat{E}$, $\hat{F}$, and $\hat{H}$ 
are the generators of the quantum algebra ${\rm U_q(sl(2))}$ with the 
following matrix elements in the representation $\pi_j$:
\begin{equation}
\label{EFH}
\begin{array}{cccc}
\pi_j(\hat{H})_{m n} = (j+1-n) \: \delta_{m,n}, 
& m,\: n = 1,\: 2, \: \ldots, \: 2j+1. \\
\pi_j(\hat{E})_{m n} = \omega_{m} \: \delta_{m, n-1}, &
\pi_j(\hat{F})_{m n} = \omega_{n} \: \delta_{m-1, n}. \\
\end{array}
\end{equation}
We introduce the notation $\omega_{n} \; = \; \sqrt{[n]_q [2j+1-n]_q}$;
$[x]_q=(q^x-q^{-x})/(q-q^{-1})$. 
The $R$-matrices with higher spins $l > \frac12$ can be relatively easily 
obtained by solving Yang-Baxter equation~(\ref{YANGBAXTER}) 
recursively. For $R_{\hat{1} \otimes 1}(u)$, for example, we obtain
\begin{eqnarray}
R_{11,11},R_{33,33} & = & \sin(u+\eta)\sin(u+2\eta) \\
\nonumber
R_{11,33},R_{33,11} & = & \sin(u)\sin(u-\eta)\\
\nonumber
R_{13,31},R_{31,13} & = & 2\sin(\eta)\sin^2(\eta) \\
\nonumber
R_{11,22},R_{33,22},R_{22,11},R_{22,33}&=& \sin(u)\sin(u+\eta) \\
\nonumber
R_{12,21},R_{21,12},R_{23,32},R_{32,23}&=& \sin(2\eta)\sin(u+\eta)\\
\nonumber
R_{12,32},R_{21,23},R_{23,21},R_{32,12}&=& \sin(2\eta)\sin(u) \\
\nonumber
\cos(\eta)-\frac12\cos(3\eta)-\frac12\cos(2u+\eta) & = & R_{22,22}.
\end{eqnarray}
We set the normalization of the $R$-matrices to be such that the matrix 
element $R_{11,11}(u)$ represents a product of the form 
$\sin(u+\eta)\sin(u+2\eta)\ldots \sin(u+2 l \eta)$  In this normalization, 
the $R$-matrix has no poles in the complex plane of the spectral 
parameter~$u$. In the chosen normalization, if we introduce new variables
$t_{l, j} (u)$ instead of $T_{l, j} (u)$ using the relations
\begin{eqnarray}
\label{RULE}
j=0 & : & 
t_{l, 0}(u)=
\prod_{k=-(l-\frac12)}^{l-\frac12}\phi(u+k\eta),\\
\nonumber
j\le l-\frac12 & : & 
t_{l, j}(u) = T_{l, j}(u) 
\prod_{k=-(l-j-\frac12)}^{l-j-\frac12}\phi(u+k\eta),\\
\nonumber
j>l-\frac12 & : & 
t_{l, j}(u) = T_{l, j}(u),
\end{eqnarray}
where
\begin{equation}
\label{RULE2}
\phi (u) \equiv \sin^N(u+\eta/2),
\end{equation}
than the functional fusion relations for the transfer matrices become
\begin{eqnarray}
\label{USUALFUSION}
\nonumber 
t_{l, \frac12}(u-(j+1/2)\eta) \; 
t_{l,j}(u) & = & 
t_{l,0}(u-(j+1)\eta) \; 
t_{l,j-\frac12}(u+\eta/2) \\ 
& + & t_{l,0}(u-j\eta) \; 
t_{l,j+\frac12}(u-\eta/2).
\end{eqnarray} 
The procedure of obtaining the fusion relations is known~\cite{BelStr,ZH, A1, A2},
in the course of this procedure, it is important to attend to normalization 
rules~(\ref{RULE}, \ref{RULE2}). We therefore omit the detailed derivation of
relation~(\ref{USUALFUSION}).

We call the operator
\begin{equation}
\label{SVYAZJ}
H_{XXZ}=\frac{\sin(\eta)\sin(2\eta)\ldots \sin(2 l \eta)}
{\sin^{2l-1}(\eta)}
\frac{d}{d u} \log T_{l, l}(u)|_{u=0}.
\end{equation}
the Hamiltonian of the anisotropic Heisenberg model for a chain of 
particles with spin $l > \frac12$ in chosen 
normalization~(\ref{RULE}, \ref{RULE2}), by analogy with Eq.~(\ref{SVYAZ}). 
This operator describes a chain of particles with spins
$l$ with the interaction only between nearest neighbors~\cite{PS}. 
We note that $T (u)$ in formula~(\ref{SVYAZ}) denotes
$T_{1/2,1/2}(u)$, where the indices are omitted for simplicity. 
Generally, not only the nearest neighbor interaction, i.e., the dipole 
interaction, but also the quadrupole interaction should be taken into
account when describing a chain of particles with spin $l > 1/2$, 
but this is beyond the scope of this work.

\paragraph{Connection of the transfer matrix with the first and second
linearly independent solutions of the Baxter $TQ$-equation.}
The transfer matrix $T_{l, j} (u)$ is equal to the trace of the monodromy 
matrix with respect to the auxiliary space~\cite{TF}. 
The monodromy matrix $L_{l \otimes j} (u)$ acts in the configuration 
space $({\bf C}^{2l+1})^{\otimes N}\otimes {\bf C}^{2j+1}$. 
In the case where $j=\frac12$, we have
\begin{equation}
L_{l \otimes \frac12} (u) = q^{-\beta \sigma^z}
R_{l \otimes \frac12}(u) \ldots 
R_{l \otimes \frac12}(u) =
\left(
\begin{array}{cc}
q^{-\beta} A_{l}(u) & q^{-\beta} B_{l}(u) \\
q^{+\beta} C_{l}(u) & q^{+\beta} D_{l}(u) \\
\end{array}
\right).
\end{equation}
Using the quantum inverse problem method, we determine the vacuum vector
$|0\rangle$ (all spins are directed upwards). Taking explicit 
expression~(\ref{R22}) for the $R$-matrix into account, we obtain
\begin{eqnarray}
C_{l}(u) |0\rangle & = & 0 \\
\nonumber
A_{l}(u) |0\rangle & = & \phi(u+l\eta) |0\rangle \\
\nonumber
D_{l}(u) |0\rangle & = & \phi(u-l\eta) |0\rangle
\end{eqnarray}
for this vector. If $\{v_1, v_2, \ldots, v_n \}$ 
are the roots of the Bethe equations (see Eq.~(\ref{BE}) below), then the Bethe
eigenvector is
\begin{equation}
\psi(\vec{v})=B_{l}(v_1)\ldots B_{l}(v_n)\;|0\rangle
\end{equation}
Using the commutation relations for $A_{l}(u)$, $B_{l}(u)$, and 
$D_{l}(u)$~\cite{TF} following from Yang-Baxter equation~(\ref{YANGBAXTER}),
we obtain the expression for the eigenvalues of the transfer 
matrix $T_{l,\frac12}(u)=q^{-\beta}A_{l}(u) + q^{\beta}D_{l}(u)$
\begin{eqnarray}
\label{SZ}
T_{l, \frac12}(u) & = &
q^{-\beta} \phi(u+ l\eta)\prod^{n}_{i=1}
\frac{\sin(u-v_i-\eta)}{\sin(u-v_i)}\\
\nonumber
&+&q^{+\beta} \phi(u-l \eta)\prod^{n}_{i=1}
\frac{\sin(u-v_i+\eta)}{\sin(u-v_i)}.
\end{eqnarray}
The Bethe equations themselves can be derived taking into account 
that $T_{l, \frac12}(u)$ has no poles in the complex plane of the 
spectral parameter $u$. The equations have the form
\begin{eqnarray}
\label{BE}
\frac{\phi(v_k+l \eta)}{\phi(v_k-l \eta)}
& = & q^{2 \beta} \prod^{n}_{i=1}
\frac{\sin(v_k-v_i+\eta)}{\sin(v_k-v_i-\eta)}.
\end{eqnarray}
The Baxter's $Q$-operator acts in the space $({\bf C}^{2l+1})^{\otimes N}$. 
By definition, it has the eigenvalues
\begin{equation}
\label{DEFQ}
Q_{l}(u) = \prod^{n}_{i=1} \sin(u - v_i).
\end{equation}
Taking Eq.~(\ref{SZ}) into account, we obtain the Baxter $TQ$-equation, 
generalized for the case of an arbitrary spin $l$:
\begin{equation}
\label{TQJ}
T_{l, \frac12}(u) Q_{l}(u) = 
q^{-\beta}\phi (u+l \eta) Q_{l} (u-\eta) +
q^{+\beta}\phi (u-l \eta) Q_{l} (u+\eta).
\end{equation}
According to~\cite{BelStr}, such an equation can be interpreted as a 
discrete analogue of a second-order differential equation; 
therefore, there must exist a second linearly independent solution 
$P_{l} (u)$ corresponding to the same eigenvalue of the transfer matrix:
\begin{equation}
\label{TPJ}
T_{l, \frac12}(u) P_{l}(u) = 
q^{-\beta}\phi (u+l \eta) P_{l} (u-\eta) +
q^{+\beta}\phi (u-l \eta) P_{l} (u+\eta).
\end{equation}
The connection between the transfer matrix and the operators $P$ and $Q$
can be determined. We do not go into the details of the calculation and 
just write the result:
\begin{eqnarray}
\label{TPQ}
t_{l, j} (u) & = & 
q^{\beta}q^{\frac{2\beta}{\eta} u}
\left[
P_{l}\left(u+(j+\frac12)\eta \right)
Q_{l}\left(u-(j+\frac12)\eta\right) \right. \\
\nonumber
&-& 
\left.
Q_{l}\left(u+(j+\frac12)\eta\right)
P_{l}\left(u-(j+\frac12)\eta\right)
\right],
\end{eqnarray}
where $t_{l, j} (u)$ is connected with $T_{l, j} (u)$ through 
relations~(\ref{RULE}). 
Formula~(\ref{TPQ}) also determines the normalization of the function
$P_{l} (u)$ not fixed by Eq.~(\ref{TPJ}). Strictly speaking, relation~(\ref{TPQ})
is a hypothesis confirmed by the coincidence of Eq.~(\ref{USUALFUSION}) with 
the functional fusion relations for the transfer matrices obtained 
using connection~(\ref{TPQ}), while formula~(\ref{TPQ}) can be easily 
verified by direct calculation if the spin values are suficiently
small. Therefore, formula~(\ref{TPQ}) can be proved by induction. 
Analytic properties of the function $Q_{l}(u)$ are
determined by formula~(\ref{DEFQ}), for example, $Q_{l}(u+\pi)=(-1)^n Q_{l}(u)$.
In the next section, we investigate the properties of the function $P_{l}(u)$
in the case where the anisotropy parameter is equal to the root of minus one.

\paragraph{Truncation of fusion relations for the anisotropy parameter
$q^{p+1}=-1$.}
It follows from relation~(\ref{TPQ}) that for $q^{p+1}=-1$ 
(i.e., $\eta=\frac{\pi}{p+1}$) and with the condition
\begin{equation}
\label{REL}
\frac{Q(u+\pi)}{Q(u)}=\frac{P(u+\pi)}{P(u)},
\end{equation}
satisfied, fusion relations~(\ref{USUALFUSION}) are truncated, i.e.,
\begin{eqnarray}
\label{TRUN}
t_{l, \frac{p}{2}} (u) & = & 0 \\
\nonumber
t_{l, \frac{p}{2}-j-\frac12} (u) & = & 
-(-1)^{n-\beta}t_{l, j}(u+\frac{\pi}{2}).
\end{eqnarray}
We investigate the conditions under which requirement~(\ref{REL})
is satisfied. We set $j = 0$ in Eq.~(\ref{TPQ}). Using Eq.~(\ref{RULE}), 
we obtain
\begin{equation}
\label{EQ25}
q^{-\beta}\frac{\prod_{k=1}^{2l}\phi(u+(k-l-\frac12)\eta)}
{Q(u-\eta/2)Q(u+\eta/2)}
=
q^{\frac{2\beta}{\eta} u}\left( 
\frac{P(u+\eta/2)}{Q(u+\eta/2)}-\frac{P(u-\eta/2)}{Q(u-\eta/2)}
\right).
\end{equation}
The left-hand side of Eq.~(\ref{EQ25}) is a ratio of trigonometric 
polynomials that can be uniquely expanded in a sum:
\begin{equation}
q^{-\beta}\frac{\prod_{k=1}^{2l}\phi(u+(k-l-\frac12)\eta)}
{Q(u-\eta/2)Q(u+\eta/2)}
=
K(u)+q^{-\beta}\frac{G(u+\frac{\eta}{2})}{Q(u+\frac{\eta}{2})}
-q^{\beta}\frac{G(u-\frac{\eta}{2})}{Q(u-\frac{\eta}{2})},
\end{equation}
where $K(u)$ and $G(u)$ are uniquely determined trigonometric polynomials 
of the degrees $\deg(K(u))=2 l N-2 n$ and $\deg(G(u))<n$. We also use 
another expansion,
\begin{equation}
\label{KF}
K(u) = q^{-\beta}F(u+\frac{\eta}{2})-q^{\beta}F(u-\frac{\eta}{2}),
\end{equation}
and then
\begin{equation}
P(u) = q^{\frac{-2\beta}{\eta} u}(Q(u)F(u)+G(u)).
\end{equation}
We require that the behavior of the function $F(u)$ under the shift of the 
argument by $\pi$ is same as that of the function $F(u)$,
i.e., $K(u+\pi)=(-1)^{2lN} K(u)$. We then obtain the condition
\begin{equation}
\label{USL1}
(-1)^{2\beta} = (-1)^{2lN}.
\end{equation}
Solving Eq.~(\ref{KF}) by expanding the functions $K(u)$ and $F(u)$
in Fourier series, we obtain yet another condition,
\begin{equation}
\label{USL2}
l N-n \; < \; \min(\beta,\; p+1-\beta) \mod (p+1).
\end{equation}
Therefore, conditions~(\ref{USL1}) and~(\ref{USL2}) are necessary for 
truncation~(\ref{TRUN}) of functional fusion relations~(\ref{USUALFUSION}).

\paragraph{Algebraic structure.}
It is known that for the $XXZ$-chain of particles whose spins are $1/2$, 
the truncation of functional relations~(\ref{USUALFUSION}) is explained
by the transfer matrix$T_{l, \frac{p}{2}} (u)$ being expressible either as
\begin{equation}
\label{TXA}
T_{l, p/2}(u) = X^p(\ldots) \; + \; (\ldots) X,
\end{equation}
in the case of open boundary conditions of special type~\cite{We} or as
\begin{equation}
\label{TXB}
T_{l, p/2}(u) = X^{p+1-\delta}(\ldots) \; + \; (\ldots) X^{\delta},
\end{equation}
in the case of generalized periodic boundary conditions~\cite{BelGub},
where $\delta=\beta-s$ is a positive integer number, the integer or half-integer 
$s$ is an eigenvalue of the operator $S^z$ ($\beta$ is integer or half-integer
as $s$, because $\delta$ always integer), and the operators $X$ and $S^z$
are the generators of the quantum algebra ${\rm U_q(sl(2))}$ with the forms
\begin{equation}
X =  \sum_{i=1}^{N} q^{\frac12(\sigma^z_{1}+\ldots+\sigma^z_{i-1})}
\sigma^{+}_{i} q^{-\frac12(\sigma^z_{i+1}+\ldots+\sigma^z_{N})}, \qquad
S^z = \sum_{i=1}^{N} \frac12\sigma^z_i.
\end{equation}
The dots $(\ldots)$ in formulas~(\ref{TXA}) and~(\ref{TXB}) denote certain
operators obtained in~\cite{We} and~\cite{BelGub}, whose explicit
form is unimportant in this analysis. The truncation of the fusion relations
occurs on the cohomologies
\begin{equation}
V_{p} = \Ker X/\im X^{p},
\end{equation}
in the case described by Eq.~(\ref{TXA}) and on the cohomologies
\begin{equation}
\label{VPL}
V_{p, \delta} = \Ker X^{\delta}/\im X^{p+1-\delta}.
\end{equation}
in the case described by Eq.~(\ref{TXB}). Practically, the particle spin 
values are not used for obtaining formulas~(\ref{TXA},~\ref{TXB}). 
Therefore, formula~(\ref{TXB}) also holds in our case, 
but with one correction: in this case, the generators
$X$ and $S^z$ of the quantum algebra ${\rm U_q(sl(2))}$ are expressed by 
the formulas
\begin{equation}
X =  \sum_{i=1}^{N} q^{\hat{H}_{1}+\ldots+\hat{H}_{i-1}}
\hat{E}_{i}q^{-\hat{H}_{i+1}-\ldots-\hat{H}_{N}}, \quad
S^z = \sum_{i=1}^{N}\hat{H}_i.
\end{equation}
In this formula, the matrix elements of the operators
$\hat{E}$, $\hat{F}$, and $\hat{H}$  are given by expression~(\ref{EFH})
where $j$ should be replaced with $l$.

\paragraph{Chain of particles with spin $l = 1$ with the anisotropy parameter
equal to root of minus one.}
We consider the simplest case where $q^5=-1$. 
We use the notation $G(u)\equiv T_{\hat{1},1}(u+2\pi/5)$ for convenience. 
We then obtain the functional equations
\begin{eqnarray}
\label{PRS}
G(u+\frac{\pi}{2})&=& G(u-\frac{\pi}{2})\\
\nonumber
G(u+\frac{\pi}{10})G(u-\frac{\pi}{10})&=&
\sin^N(u+\frac{\pi}{10})\sin^N(u-\frac{\pi}{10})\\
\nonumber
&\times& 
\left( G(u+\frac{\pi}{2})+ 
\sin^N(u+\frac{3\pi}{10})\sin^N(u-\frac{3\pi}{10}) \right),
\end{eqnarray}
from relations~(\ref{RULE}, \ref{TRUN}).
It can be seen that $G(-u) = G(u)$. Because the transfer-matrix eigenvalues
have no poles with respect to $u$, the solution of Eqs.~(\ref{PRS})
should be sought in the form
\begin{equation}
G(u) = \sum_{k=0}^{N} g_{k} \cos(2 k u).
\end{equation}
We now consider the chain of four particles. In this case, the solution of 
system~(\ref{PRS}) has the form
\begin{eqnarray}
G_{1,2}(u) & = & \frac{1}{32}\sin^4(u)\left(
6 \pm 5(\sqrt{5}-3)+5(\sqrt{5}-1)\cos(2u) + \right. \\
\nonumber
 &  & + \left. 2(\sqrt{5}+1)\cos(4u) \right), \\
\nonumber
G_3(u) &=& \frac{1}{32}\sin^4(u)\left(
3\sqrt{5}-9+4(\sqrt{5}-2)\cos(2u)- \right. \\
\nonumber
& & \left. -2(\sqrt{5}-1)\cos(4u) \right),\\
\nonumber
G_{4,5}(u) & = & \sin^4(u)\left(
\frac{37-15\sqrt{5}\pm\sqrt{430-70\sqrt{5}}}{64} + \right. \\
\nonumber
&  & + \left. \frac{\sqrt{5}-4\pm\sqrt{205-80\sqrt{5}}}{16}\cos(2u)
-\frac{\sqrt{5}-1}{16}\cos(4u) \right).
\end{eqnarray}
Using formula~(\ref{SVYAZJ}), we obtain the corresponding energy levels:
\begin{eqnarray}
\label{EGG}
E_{1,2} & = & (1-\sqrt{5})\pm (1+\sqrt{5}),\\
\nonumber
E_3 & = & -\frac{1+3\sqrt{5}}{2},\\
\nonumber
E_{4,5} & = & \frac14 \left( 7-3\sqrt{5} \pm \sqrt{94+22\sqrt{5}}.
\right) 
\end{eqnarray}
To verify this result, we calculated the energy eigenvalues and compared 
them with those given by Eqs.~(\ref{EGG}).
Levels~(\ref{EGG}) are unique, while the other levels in the same sectors
are degenerate.

\paragraph{Conclusion.}
We have proved that the fusion relations for transfer matrices with an 
arbitrary spin $l$ are truncated. The truncation occurs if 
conditions~(\ref{USL1},~\ref{USL2}) are satisfied on the space of 
cohomologies~(\ref{VPL}). The truncation results in a closed system 
of equations whose solution determines the energy spectrum of states
in the reduced Heisenberg $XXZ$-model for a chain of particles whose 
spin is $l$ with generalized periodic boundary conditions.

\paragraph{Acknowledgments.} 
The author thanks A. A. Belavin and M. Yu. Lashkevich for the useful comments
and fruitful discussions. This work is partially supported by RFBR 01-02-16686
and 00-15-96579, CRDF RP1-2254 and INTAS-00-00055.

\end{document}